\documentclass[pdflatex,sn-mathphys]{sn-jnl}
\usepackage{amsmath, listings, colortbl, caption, subcaption, multirow}

\theoremstyle{thmstyleone}%
\theoremstyle{thmstyletwo}%

\theoremstyle{thmstylethree}%

\begin{document}

\title[A Genetic Algorithm for Hadamard Matrices on GPUs]{A GPU accelerated Genetic Algorithm for the Construction of Hadamard Matrices}

\author*[1]{\fnm{Andras} \sur{Balogh}}\email{andras.balogh@gutrgv.edu}

\author[1]{\fnm{Raven} \sur{Ruiz}}\email{ruiziraven@gmail.com}

\affil[1]{\orgdiv{School of Mathematical and Statistical Sciences}, \orgname{The University of Texas Rio Grande Valley}, \orgaddress{\street{1201 West University Drive}, \city{Edinburg}, \postcode{78539}, \state{TX}, \country{USA}}}

\abstract{
We use a genetic algorithm to construct  Hadamard Matrices. The initial population of random matrices is generated to have a balanced number of $+1$ and $-1$ entries in each column except the first column with all $+1$. Several fitness functions are implemented in order to find the most effective one. The crossover process creates offspring matrix population by exchanging columns of the parent matrix population. The mutation process flips $+1$ and $-1$ entry pairs in random columns. The use of CuPy library in Python on graphics processing units enables us to handle populations of thousands of matrices and matrix operations in parallel.  
}


\maketitle

\section{Introduction}\label{section:intro}
Hadamard matrices are square matrices with $+1$ and $-1$ entries where the columns are mutually orthogonal.
According to the Hadamard Conjecture \cite{hadamard},   Hadamard matrix of size $m\times m$ (order $m$) exists when $m=1$, $m=2$, and when $m=4k$ for arbitrary $k$ positive integer.  Deterministic algorithms are known for creating Hadamard matrices of order $2^{k}$ and only a few other special cases of order $4k$.  In \cite{Goethals}, Goethals and Siedel has shown  that there exist skew Hadamard matrices of order $m=36$ and $m=52$. The Goethals--Siedel method was implemented to construct $32$ in--equivalent Hadamard matrices of order $m=404$ in \cite{Jayathilake}. A classification of Hadamard matrices using types of quadruples of rows with two distinct values were done in \cite{Mohammadian}.  In \cite{Seberry} Seberry and Yamada highlights important Hadamard matrix theorems, use several methods to construct Hadamard matrices, display results, and analyze their findings from each method. More recently Suksmono in \cite{suksmono} applied a Simulated Annealing Algorithm and  in \cite{SuksmonoQuantum} a  Williamson based quantum computing method to construct Hadamard matrices. The reader is referred to \cite{Browne} for  a recent list of known constructions results for  Hadamard matrices of different orders.

In this work we develop a genetic algorithm for the construction of Hadamard matrices. The parallel numerical code uses CUDA GPUs to accelerate the computations. 

In Section \ref{section:GPU} we briefly describe the difference between traditional CPU computing and our approach of using GPUs to accelerate computations.

In Section \ref{section:GAH} we describe how our parallel implementation of a Genetic Algorithm works for finding Hadamard matrices. In Section \ref{section:comp-results} we discuss the computational results found by the use of our Genetic Algorithm. Finally, the paper concludes with a summary in Section \ref{conclusion}, and the numerical code in Section \ref{code}.   

\section{Computing on Graphics Processing Units}\label{section:GPU}
Stochastic algorithms and matrix calculations involve large number of calculations. A CPU (Central Processing Unit) does calculations mostly in serial way, which can take a large amount of time if working with many large matrices. GPUs (Graphics Processing Units) were developed for fast graphics rendering by calculating what to do and when with the millions of pixels in a computer screen. Since a GPU can do thousands of calculations simultaneously (in parallel), the time working with large number of matrices can be reduced significantly.  The use of GPUs  allows us to do parallel calculations more effectively than using CPUs. Most of our calculations have been performed on an EVGA GeForce RTX 3080 XC3 black  \cite{EVGA}, having 8704 CUDA cores, 10 GB memory, and CUDA capability 8.6.  With the use of GPUs our genetic algorithm was able to handle for example a population of $40,000$ matrices of size $20\times 20$. 

NumPy \cite{numpy} is a popular library for the Python programming language. It supports overloaded mathematical functions operation on multi-dimensional arrays. For example, if $A$ is a $100\times 100$ matrix defined as \lstinline|A=numpy.random.rand(100,100)|, then \lstinline|B=numpy.sin(A)| produces a matrix containing the sine of the elements of matrix $A$. The calculations of the $10,000$ elements are done one-by-one, in serial fashion on the CPU.
For this project, we are using the CuPy \cite{cupy},  Python library accelerated with NVIDIA CUDA \cite{cuda} for GPUs. It was created specifically to be highly compatible with  NumPy.  For example modifying the previously mentioned NumPy matrix example to \lstinline|A=cupy.random.rand(100,100)| and \lstinline|B=cupy.sin(A)|, the calculations are distributed on the thousands of CUDA threads to be done in parallel.  

Using the CuPy library and in general CUDA on the GPU involves the following steps. It is important to note that the CPU controls the GPU too. 

\begin{enumerate}
    \item Data is copied from CPU memory to GPU memory.
    \item CPU initiates the calculations on the GPU.
    \item GPU executes the calculations.
    \item Results are copied back from GPU to CPU.
\end{enumerate}    

Each of these steps require time. The speed up of the GPU execution has to be significant in order to balance the extra time required by copying the data back and forth between CPU and GPU. 

We include here a simple speed comparison of using Numpy (serial calculations) vs. CuPy (parallel calculations). Since our actual code with Genetic Algorithm includes raw kernels, we never wrote a serial version of it, but due to the complicated nature of the calculations with large multi-dimensional arrays, we expect the parallel computations to be thousands of times faster than the serial code. The example creates a list of $N=10,000$ random matrices of size $100\times 100$, and then multiplies each with their transform, repeating the calculation $M$ times.  

\begin{figure}
    \centering
    \caption{Cupy vs. Numpy Execution Times (seconds)}
    \label{fig:Cupy-Numpy}
    \includegraphics[width=0.65\textwidth]{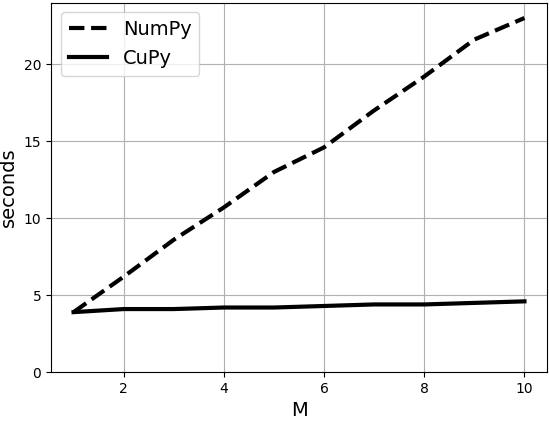}
\end{figure}

NumPy version:
\begin{lstlisting}[language=Python, numbers=none] 
import numpy as np
m=100
N=10**4
Pop = np.random.rand(N, m, m)
for i in range(M):
    F=np.matmul(np.transpose(Pop,axes=(0,2,1)),Pop)
\end{lstlisting}

CuPy version:
\begin{lstlisting}[language=Python, numbers=none] 
import cupy as cp
m=100
N=10**4
Pop=cp.random.rand(N, m, m)
for i in range(M):
    F=cp.matmul(cp.transpose(Pop,axes=(0,2,1)),Pop)
\end{lstlisting}

Note the identical syntax other than the use of "cp" for CuPy vs. "np" for NumPy. The time required for the calculations are summarized in Table \ref{tab:Cupy-Numpy} and in Figure \ref{fig:Cupy-Numpy}. For $M=1$, when the matrix calculations are done only once, the CuPy and NumPy calculations take the same $3.9$ seconds. Although the GPU does the calculation faster than the CPU, the time it takes to load the GPU slows the completion of the code down. For repeated calculations of the batched matrix multiplications the Numpy code execution time increases by about $2.12$ seconds for each additional loop steps taken. The CuPy calculation only increases by about $0.07$ seconds for each loop steps. This means that the CuPy operation is $30$ times faster than the NumPy operation after ignoring the initial data copy between the CPU and GPU.

\begin{table}
\caption{CuPy vs. NumPy Execution Times (seconds)}
    \centering
    \begin{tabular}{|l|c|c|} \hline
    M    & NumPy & CuPy \\ \hline\hline
    $1$  & $3.9$ & $3.862$ \\ \hline
    $2$  & $6.2$ & $3.994$ \\ \hline
    $3$  & $8.6$ & $4.038$ \\ \hline
    $4$  & $10.7$ & $4.095$ \\ \hline
    $5$  & $13.0$ & $4.175$ \\ \hline
    $6$  & $14.6$ & $4.252$ \\ \hline
    $7$  & $17.0$ & $4.287$ \\ \hline
    $8$  & $19.2$ & $4.330$ \\ \hline
    $9$  & $21.6$ & $4.394$ \\ \hline
    $10$  & $23.0$ & $4.454$ \\ \hline
    $\vdots$  & $\vdots$ & $\vdots$ \\ \hline
    $100$  & $218.6$ & $9.916$ \\ \hline
    \end{tabular}
    \label{tab:Cupy-Numpy}
\end{table}

It is also possible to combine CuPy code with CUDA kernel function that use C++ syntax. We use several so-called "raw kernels" when we need more complicated operations than simple matrix-vector ones. This requires to know how threads are arranged in grids of blocks, as explained in the next section.

\section{Genetic Algorithm and Hadamard Matrices} \label{section:GAH}

Genetic algorithms are a search heuristic that was inspired by the Natural Selection process \cite{wirsansky2020pythonGA}. Natural Selection is a process where individuals adapt and change based on the environment and other variables. Each individual in the population is unique, meaning that each individual has different traits. Some individuals may have better traits than others, this allows those with better traits to live longer and reproduce. Those superior traits are then passed down to the next generation, with some variation, this is known as evolution. Evolution is a key component of genetic algorithms, it allows the algorithms to search for the optimal solution of several problems in mathematics. We will continue to discuss evolution in the following section.

Genetic algorithms use Darwin's evolution theory of natural selection as a bases to search for optimal solutions. There are three main points behind Darwin's theory of evolution that are given as follows
\begin{enumerate}
    \item \emph{Inheritance}: Each individual inherits traits that were passed down from their parents. The traits that are most likely to be passed down are those that will improve chances of survival.
    \item \emph{Variation}: Each individual in a population will have variation that is unique to them, even those that are related to each other.
    \item \emph{Fitness and Selection}: The most fit individuals are more likely to survive, the surviving individuals are then able to reproduce and pass on their genes to future generations.
\end{enumerate}
Ultimately, Darwin's evolution theory suggests that the individuals with the best traits will survive and maintain the population with offspring of their own. The offspring will most likely be better equipped for survival and each generation there after will become more adaptive. This leads to a genetic algorithm that will be used throughout this paper.

\begin{enumerate}
    \item \emph{Generate Initial Population}: Create set of individuals in a population.
    \item \emph{Compute Fitness}: A numerical measure of how close an individual is to becoming `fit'.
    \item \emph{REPEAT}
    \begin{enumerate}
        \item \emph{Selection}: Select the most fit individuals which is based on the fitness.
        \item \emph{Crossover}: The selected individuals become parents, the parents are paired and create a pair of offspring that inherit traits from the parents.
        \item \emph{Mutation}: Each offspring will have some form of variation to them.
        \item \emph{Compute Fitness}: A numerical measure of how close an individual is to becoming `fit'.
    \end{enumerate}
    \item \emph{UNTIL Population has converged}: An individual has met the required fitness.
\end{enumerate}

The fundamental theorem of genetic algorithms \cite{BridgesG87} is also know as Holland's schema theorem \cite{Holland}, proposed by John Holland in the 1970's. While the mathematics is quite involved, we will focus on the basic understanding of the theorem. The theorem suggests that an individual will prevail if it has an above average fitness.

In Subsection \ref{subsection:Population}, we detail how an initial population of matrices is generated. For this process, two methods were implemented and we discuss their differences. In Subsection \ref{subsection:Fiteness-Func}, we use a fitness function to compute the fitness of each matrix in a population. These fitness values are used as a measurement to see how close a matrix is to becoming an Hadamard matrix. There are four fitness functions that were implemented and we discuss their differences. In Subsection \ref{subsection:Selection}, we use the fitness values of each matrix to select parent matrices from a population. In Subsection \ref{subsection:Crossover}, we use a process that creates offspring matrices from selected parent matrices. In Subsection \ref{subsection:Mutation}, we discuss how each offspring matrix is given some variation or mutation. Three methods were implemented that achieve this.
\subsection{Population} \label{subsection:Population}
The population consists of $4N$ matrices, each of them of size $m \times m = 4k \times 4k$, where $k,N\in\mathbb{N}$. They are arranged in the form of a 3-dimensional array:
\begin{align} \label{eq:Population}
    \text{Population} & = \left[Q_{1},\ldots,Q_{4N}\right]\\
    & =\left[P_{1},\ldots,P_{N},P_{N+1},\ldots,P_{2N}, O_{1},\ldots,O_{N},O_{N+1},\ldots,O_{2N}\right],
\end{align}
with $P_{1},\ldots,P_{2N}$ denoting the parent matrices and $O_{1},\ldots,O_{2N}$ denoting the offspring matrices.  During the crossover process matrices $P_1,\ldots,P_N$ are paired up with matrices $P_{N+1},\ldots,P_{2N}$  to create  offspring matrices $O_1,\ldots,O_{2N}$. 

At the beginning of the process the population of matrices are initialized  to have a balanced number of $+1$ and $-1$ entries in each column except for the first column that consists of all $+1$ entries. This way columns $2--m$ are automatically orthogonal to the first column, and the first columns will not have to be changed during the process. 
In order to achieve a random order of the  $+1$ and $-1$ entries in columns $2-m$ of each matrix, first all matrices are initialized to have all $+1$ entries, then the upper half is filled with $-1$ (see code lines \ref{line:popinit1}--\ref{line:popinit2}). Then we implement a raw (CUDA) kernel function called \lstinline|Shuffle|\_\lstinline|Column| that uses the Fisher-Yates shuffle  algorithm \cite{Fisher:1938:STB, Durstenfeld} to shuffle the $\pm 1$ entries in columns $2-m$ of each matrix.  While the shuffling in each column is done in serial, the columns are handled simultaneously in parallel. See code lines \ref{line:shuffle_begin}--\ref{line:shuffle_end} for the \lstinline|Shuffle|\_\lstinline|Column| raw kernel, code line
\ref{line:popinitseed} for generating a random seed, and code line \ref{line:popinitcall} for calling the raw kernel.


\subsection{Fitness Function} \label{subsection:Fiteness-Func}
In the natural selection process, the most fit members of the population will most likely to survive and reproduce. For our genetic algorithm, the fitness function measures how close a matrix is to being an Hadamard matrix. For an $m\times m$ matrix $Q$ with $\pm 1$ entries the product matrix $Q^TQ$ has diagonal entries $m$ and this product matrix is diagonal if and only if $Q$ is Hadamard matrix. Also, the closer the off-diagonal entries are to zero, the closer the corresponding matrix columns are to orthogonal. This leads to the fitness function
\begin{equation}
    F_1 = \sum_{i,j=1}^m \left\vert  \left[Q^{T}Q\right]_{ij} \right\vert -m^{2}. \label{fitness0}
\end{equation}
Instead of fitness function \eqref{fitness0}, we chose the slightly more computationally efficient fitness function
\begin{equation}
f=\text{number of nonzero entries}\left(Q^{T}Q\right)-m,  \label{eq:fitness}
\end{equation}
which is also a nonnegative function with a zero minimum value attained for Hadamard matrices only. The fitness values are calculated for each matrix in the population resulting in an array
\begin{equation} \label{eq:Fitness-Form}
    F=\left[f_{1},\ldots,f_{N},f_{N+1},\ldots,f_{2N}, f_{2N+1},\ldots,f_{3N},f_{3N+1},\ldots,f_{4N}\right],
\end{equation}
where $f_{1},\ldots,f_{2N}$ and $f_{2N+1},\ldots,f_{4N}$ refers to the fitness of $P_{1},\ldots,P_{2N}$ and $O_{1},\ldots,O_{2N}$ in the population  respectively. The calculation of fitness values \eqref{eq:fitness} implemented using CuPy can be found on code lines  \ref{line:fitness1} and \ref{line:fitness2}.

\subsection{Selection} \label{subsection:Selection}
The purpose of the selection process is to select matrices that are more likely to become Hadamard matrices based on their fitness value in $\eqref{eq:Fitness-Form}$. The selected matrices will take the parent position  (first half) of the population in \eqref{eq:Population} in random order for future use while the rest of the matrices are discarded.  The compact CuPy implementation of the selection process is on code line  \ref{line:selection}. 

The literature often uses selection process with probabilities proportional to how fit an individual is. In our case we want to minimize the fitness function  and hence the probabilities would have to be inverse proportional to the fitness values. We were not able to find probabilistic selection process that would be more effective than the the above deterministic one.  

\subsection{Crossover} \label{subsection:Crossover}
The purpose of the crossover is to create a pair of offspring matrices from a pair of parent matrices. The parent matrices $\left[ P_1,\ldots,P_N,P_{N+1},\ldots,P_{2N} \right] $ are paired up as  $\left(P_1, P_{N+1}\right)$, $\left(P_2, P_{N+2}\right)$, \ldots, $\left(P_N, P_{2N}\right)$. For each pair of matrices a random column index $1<c_{cop}<m$ is generated as crossover point. This crossover point splits the columns of parent matrices in each pair into two parts: columns $1,\ldots,c_{cop}$ and columns $c_{cop}+1,\ldots, m$.  Each part in one matrix is matched with the adjacent  part from the other matrix to create  a pair of offspring matrices.   Figure \ref{fig:crossover} shows an example with a pair of $4 \times 4$ parent matrices $(P_1,P_{N+1})$, with a crossover point \lstinline|c_cop = 3| and with resulting offspring matrices $(O_1,O_{N+1})$.

\begin{figure}
    \centering
    \caption{Crossover}
    \label{fig:crossover}
\begin{align*}
    P_{1}=\left(\begin{array}{>{\columncolor{yellow!50}}c >{\columncolor{yellow!50}}c >{\columncolor{yellow!50}}c >{\color{red}}c >{\columncolor{orange!35}}c}
        -&+&+&\vert &+ \\
        -&+&-&\vert &- \\
        +&-&+&\vert &- \\
        +&-&-&\vert &+
    \end{array}\right), & \quad
    P_{N+1}=\left(\begin{array}{>{\columncolor{blue!30}}c >{\columncolor{blue!30}}c >{\columncolor{blue!30}}c >{\color{red}}c >{\columncolor{green!30}}c}
                    -&+&+&\vert &+ \\
                    +&-&+&\vert &+ \\
                    -&-&-&\vert &- \\
                    +&+&-&\vert &-
    \end{array}\right) \\[4ex]
    O_{1}=\left(\begin{array}{>{\columncolor{yellow!50}}c >{\columncolor{yellow!50}}c >{\columncolor{yellow!50}}c >{\color{red}}c >{\columncolor{green!30}}c}
                    -&+&+&\vert &+ \\
                    -&+&-&\vert &+ \\
                    +&-&+&\vert &- \\
                    +&-&-&\vert &-
    \end{array}\right), & \quad
    O_{N+1}=\left(\begin{array}{>{\columncolor{blue!30}}c >{\columncolor{blue!30}}c >{\columncolor{blue!30}}c >{\color{red}}c >{\columncolor{orange!35}}c}
                    -&+&+&\vert &+ \\
                    +&-&+&\vert &- \\
                    -&-&-&\vert &- \\
                    +&+&-&\vert &+
    \end{array}\right)    
\end{align*}
\end{figure}

The calling of the crossover process is from code lines \ref{line:crossoverpoint}-\ref{line:crossover} with the raw CUDA kernel function \lstinline|Crossover(grids, blocks, (Pop, c_cop, m, N))| on code lines \ref{line:crossover_begin}-\ref{line:crossover_end}. The \lstinline|Crossover|  function creates $2N$ offspring matrices using $2N$ parent matrices and it returns the size $4N$ array \lstinline|Pop| of matrices such that 
\begin{equation}
    \text{\lstinline|Pop|}=\left[P_{1},\ldots,P_{N},P_{N+1},\ldots, P_{2N},O_{1},\ldots,O_{N},O_{N+1},\ldots, O_{2N}\right].
\end{equation} 
The crossover process is parallelized for the pairs of parent matrices indexed by $k$ and for their rows and columns indexed by $i$ and $j$ respectively with the array of matrices  flattened. The $(i,j)$ entry of the $k^{\text{th}}$ matrix is referenced as \lstinline|Q[m*m*k+m*i+j]|.

\subsection{Mutation} \label{subsection:Mutation}
In this next section, we  discuss the mutation process in the matrix population.
The purpose of the mutation process is for each offspring matrix to have some kind of variation to them. The mutation of the matrices is done by switching one or more pairs of $\pm 1$ entries while maintaining the balance of $\pm 1$ entries in each column. We implemented three methods to achieve variation among offspring matrices:
\begin{enumerate}
    \item Flipping the same pair of  $\pm 1$ entries in each offspring matrix, by using the same random column index and same two random row indices using CuPy code 
    \begin{lstlisting}[numbers=none]
colindx = np.random.randint(1,m)
rowindx1 = np.random.randint(m)
rowindx2 = np.random.randint(m)
    \end{lstlisting}
    \item Flipping a different pair of $\pm 1$ entries in different offspring matrices by randomly selecting columns and  pairs of rows indices for each matrix using CuPy code
    \begin{lstlisting}[numbers=none]
colindx = cp.random.random_integers(1,m-1,size=(2*N))
rowindx1 = cp.random.random_integers(0,m-1,size=(2*N))
rowindx2 = cp.random.random_integers(0,m-1,size=(2*N))
\end{lstlisting}
    \item  Flipping several pairs of $\pm 1$ entries in several ($NC$) random columns and several ($NR$) random pairs of rows in each offspring matrix.
    \begin{lstlisting}[numbers=none]
NC=3 #NC should be less than m
NR=2 #NR should be less than m/2
colindx = cp.random.random_integers(1,m-1,size=(2*N,NC))
rowindx1 = cp.random.random_integers(0,m-1,size=(2*N,NR))
rowindx2 = cp.random.random_integers(0,m-1,size=(2*N,NR))
\end{lstlisting}
\end{enumerate}
The third case provides the most variation, and it coincides with case two for $NC=NR=1$. In all three cases no switch is done if the numbers in the randomly selected pairs have the same sign. The first column has all $+1$ entries, and for that reason the first column is never chosen for mutation. The third form of mutations is coded on lines \ref{line:mutation1}-\ref{line:mutation2} with the actual CUDA raw kernel function on lines \ref{line:mutation_begin}-\ref{line:mutation_end}.

To demonstrate how the third mutation process works lets consider an example. Let $m=8$ and $N=2$ such that the we have a population of $8$ matrices that have sizes $8 \times 8$ with a balanced number of $\pm 1$ entries in each column. This implies there are $4$ offspring matrices that will be mutated. Let $NC=3$ be the number of columns mutated and $NR=2$ be the number of row pairs mutated such that we have the randomly generated column and row indices given by
\begin{equation} \label{eq:Example-Third-Mutation}
    \text{\lstinline|rowindx1|}=
    \begin{bmatrix}
        6 & 5 \\
        4 & 2 \\
        0 & 6 \\
        1 & 6 
    \end{bmatrix}, \quad
    \text{\lstinline|rowindx2|}=
    \begin{bmatrix}
        3 & 1 \\
        7 & 0 \\
        1 & 2 \\
        2 & 5 
    \end{bmatrix}, \quad
    \text{\lstinline|colindx|}=
    \begin{bmatrix}
        6 & 5 & 2  \\
        5 & 2 & 4 \\
        3 & 1 & 6 \\
        2 & 7 & 6
    \end{bmatrix}
\end{equation} 
The mutation of  four  matrices are given in Figure \ref{figure:Example-Third-Mutation} such that
\begin{itemize}
    \item In part (a), the indices are $\text{\lstinline|rowindx1|}=[6, \; 5]$, $\text{\lstinline|rowindx2|}=[3, \; 1]$, and $\text{\lstinline|colindx|}=[6, \; 5, \; 2]$.
    \item In part (b), the indicies are $\text{\lstinline|rowindx1|}=[4, \; 2]$, $\text{\lstinline|rowindx2|}=[7, \; 0]$, and $\text{\lstinline|colindx|}=[5, \; 2, \; 4]$.
    \item In part (c), the indicies are $\text{\lstinline|rowindx1|}=[0, \; 6]$, $\text{\lstinline|rowindx2|}=[1, \; 2]$, and $\text{\lstinline|colindx|}=[3, \; 1, \; 6]$.
    \item In part (d), the indicies are $\text{\lstinline|rowindx1|}=[1, \; 6]$, $\text{\lstinline|rowindx2|}=[2, \; 5]$, and $\text{\lstinline|colindx|}=[2, \; 7, \; 6]$.
\end{itemize}
In each column, a successful mutation is represented in green and blue (if there is more than one successful), an unsuccessful mutation is represented in red and magenta (if there is more than one unsuccessful).  We can see in Figure \ref{figure:Example-Third-Mutation} that although we do choose the number of columns and pair of rows we want to mutate, it is not guaranteed to mutate all of them. In part (a), we saw the \lstinline|colindx=2| have no mutation, the \lstinline|colindx=5| have one mutation, and the \lstinline|colindx=6| have two mutations. So, the maximum number of mutations for each column is $NR$. 

This third mutation method improves upon the issue of working with larger matrices, we have achieved greater variation among offspring matrices and obtained faster convergence to  Hadamard matrices. Using this method, the largest Hadamard matrix found so far is of the size $32 \times 32$. This is an improvement from the other two mutation methods that produced Hadamard matrices up to size $12\times 12$ only.

\begin{figure}
\renewcommand*{\arraystretch}{0.5}
    \centering
    \caption{Example of  Mutation }
    \label{figure:Example-Third-Mutation}
    \begin{subfigure}{0.45\textwidth}
        \(
        \begin{pmatrix}
            + & + & - & + & + & + & + & - \\
            + & + & \cellcolor{magenta}- & - & - & \cellcolor{green}- & \cellcolor{blue!60}+ & - \\
            + & - & + & + & + & - & - & - \\
            + & + & \cellcolor{magenta}- & - & - & \cellcolor{green}+ & \cellcolor{blue!60}- & + \\
            + & + & - & + & - & + & + & + \\
            + & - & \cellcolor{red!60}+ & - & + & \cellcolor{red!60}- & \cellcolor{green}+ & + \\
            + & - & \cellcolor{red!60}+ & - & - & \cellcolor{red!60}- & \cellcolor{green}- & + \\
            + & - & + & + & + & + & - & -
        \end{pmatrix}
        \) 
        \caption{First Matrix}
    \end{subfigure} \hfill 
    \begin{subfigure}{0.45\textwidth}
        \(
        \begin{pmatrix}
            + & + & \cellcolor{magenta}+ & - & \cellcolor{blue!60}+ & \cellcolor{green}- & + & - \\
            + & - & + & + & - & - & + & - \\
            + & + & \cellcolor{magenta}+ & + & \cellcolor{blue!60}- & \cellcolor{green}+ & + & - \\
            + & - & - & + & + & + & - & + \\
            + & + & \cellcolor{red!60}- & - & \cellcolor{green}+ & \cellcolor{red!60}+ & - & + \\
            + & - & + & + & - & - & - & + \\
            + & + & - & - & + & - & - & - \\
            + & - & \cellcolor{red!60}- & - & \cellcolor{green}- & \cellcolor{red!60}+ & + & +
        \end{pmatrix}
        \)
        \caption{Second Matrix}
    \end{subfigure} \\
    \begin{subfigure}{0.45\textwidth}
        \(
        \begin{pmatrix}
            + & \cellcolor{green}- & + & \cellcolor{red!60}- & + & - & \cellcolor{red!60}+ & - \\
            + & \cellcolor{green}+ & - & \cellcolor{red!60}- & + & - & \cellcolor{red!60}+ & + \\
            + & \cellcolor{red!60}+ & + & \cellcolor{green}- & - & - & \cellcolor{magenta}- & - \\
            + & + & - & + & - & - & - & + \\
            + & - & - & + & - & + & - & - \\
            + & - & - & + & - & + & + & + \\
            + & \cellcolor{red!60}+ & + & \cellcolor{green}+ & + & + & \cellcolor{magenta}- & + \\
            + & - & + & - & + & + & + & -
        \end{pmatrix}
        \)
        \caption{Third Matrix}
    \end{subfigure} \hfill
    \begin{subfigure}{0.45\textwidth}
        \(
        \begin{pmatrix}
            + & - & - & - & - & - & + & - \\
            + & + & \cellcolor{red!60}+ & + & + & - & \cellcolor{red!60}+ & \cellcolor{green}+ \\
            + & + & \cellcolor{red!60}+ & + & + & - & \cellcolor{red!60}+ & \cellcolor{green}- \\
            + & - & + & - & + & + & - & + \\
            + & - & - & + & - & - & - & + \\
            + & + & \cellcolor{green}- & - & - & + & \cellcolor{green}+ & \cellcolor{blue!60}- \\
            + & - & \cellcolor{green}+ & + & + & + & \cellcolor{green}- & \cellcolor{blue!60}+ \\
            + & + & - & - & - & + & - & -
        \end{pmatrix}
        \)
        \caption{Fourth Matrix}
    \end{subfigure}
\end{figure}

\vspace{3cm}

\section{Computational Results} \label{section:comp-results}
In this section we further comment on the efficiency of our CUDA accelerated code and discuss the computational results.

\subsection{Local Minimum} \label{subsection:Local-Min}
\begin{figure}
    \centering
    \caption{Stalled convergence of the Fit function}
    \label{fig:fit}    \includegraphics[width=0.5\textwidth]{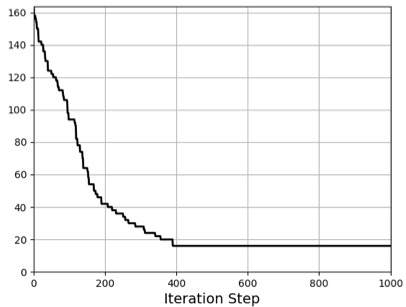}
\end{figure}

Genetic Algorithms also suffer from finding local minimums instead of global minimum similarly to the Simulated Annealing Algorithm \cite{suksmono, wirsansky2020pythonGA}. Figure \ref{fig:fit} shows the minimum of the fit function as a function of the iteration number for matrix size $20\times 20$ with $NC=NR=4$.  It is not just that the minimum of the fit function gets stuck at a positive constant without converging to zero, but the whole parent population becomes homogeneous. 
One way to try to prevent stalling at a local minimum is to use a selection process with some probability. This approach did not result in improved convergence. We had some success to speed up convergence by selecting low number of columns for the mutations. This is discussed in Subsection \ref{subsection:Mutation-Comparisons}. 
\subsection{Fitness Function Comparisons} \label{subsection:Fitness-Func-Compare}
We  compare the following two different  fitness functions:
\begin{align}
    F_{1} & =\sum_{i,j}\left\vert Q^{T}Q\right\vert-m^{2} \geq 0 \\
    F_{2}&=\text{nonzero}\left(Q^{T}Q\right)-m \geq 0 \label{fitness2}
\end{align}

Table \ref{Table:Fitness-Comparisons2} shows the average speed (in seconds) for the two fitness functions while working with $40,000$ matrices at a time and over $10,000$ iterations.  The execution time was averaged through $10$ runs. The parameter are $N=10^{4}$, $T=10^{4}$, $NC=4$, and $NR=2$ for several matrix sizes. The purpose of this table was to compare the speed of each fitness function while working with large number of matrices. Note that none of these runs resulted in any Hadamard matrices due to the relatively small number of interations.

From Table  \ref{Table:Fitness-Comparisons2} we can see that $F_{2}$ is the superior fitness function. These results were showed for only $10^{4}$ iterations. Typically when finding larger Hadamard matrices, $10^{7}$ iterations and greater is usually needed to find them. Consequently, for a larger number of iterations, $F_{2}$ is expected to perform significantly better. For these reasons, for the results in Section \ref{subsection:Results}, we use the fitness function $F_{2}$ given in \eqref{fitness2}.

\begin{table}
\caption{Speed of code completion in seconds  using $40000$ matrices and $10000$ iterations}
    \centering
    \begin{tabular}{|c|c|c|} \hline
    Matrix sizes     & $F_{1}$  & $F_{2}$  \\ \hline\hline
    $20 \times 20$   & $60.50$  & $59.38$   \\ \hline
    $40 \times 40$   & $103.15$ & $99.84$ \\ \hline
    $60 \times 60$   & $176.28$ & $170.20$\\ \hline
    $80 \times 80$   & $275.92$ & $256.59$\\ \hline
    $100 \times 100$ & $416.68$ & $381.38$ \\ \hline
    \end{tabular}
    \label{Table:Fitness-Comparisons2}
\end{table}
\subsection{Mutation Comparisons}  \label{subsection:Mutation-Comparisons}
Table \ref{Table:Average-Iterations-12} shows the average number of iterations required to find a $12 \times 12$ Hadamard matrix given different $NC$ and $NR$ values. For each case $10$ runs were made. The numbers in the table suggests that mutating two pairs ($NR=2$) in one column ($NC=1$) takes the least amount of steps to find an Hadamard matrix. In addition, the best results occurred when $NC=1$ with any amount of row pairs.

Table \ref{Table:Average-Iterations-16} shows the average number of iterations required to find a $16 \times 16$ Hadamard matrix given different $NC$ and $NR$ values. For each case $10$ runs were made. We found that the smallest numbers for $NR=1$ and $NC=2,3,4$ were obtained with the majority runs not finding Hadamard matrix. This suggest that either an Hadamard matrix was found fast or it wasn't found at all. The best results occurred when $NC=1$ with any amount of row pairs. Although it took slightly more iterations to find the Hadamard matrices, it finding them more consistently. For the case of larger matrices, this table summarizes those results. Therefore when finding Hadamard matrices, we use a low number of columns $(NC=1)$ or $(NC=2)$ and a larger number of row pairs depending on the size of the matrix. 

\begin{table}
\caption{Average Iteration steps for a $12 \times 12$ matrix using the Mutation kernel function}
    \centering
   \begin{tabular}{c||cccc} 
    & \multicolumn{4}{c}{NC} \\ 
    NR  & $1$ & $2$ & $3$ & $4$ \\ \hline \hline  $1$  & $34.2$ & $35.8$ & $37.4$ & $35.8$ \\ \hline $2$  & $32.3$ & $38.8$ & $40.5$ & $43.4$  \\ \hline     $3$  & $35.2$ & $42.6$ & $47.5$ & $50.7$  \\ \hline $4$  & $36.3$ & $46.0$ & $49.9$ & $65.7$  \\ \hline
    \end{tabular}
    \label{Table:Average-Iterations-12}
\end{table}

\begin{table}
\caption{Average Iteration steps for a $16 \times 16$ matrix using the Mutation kernel function}
    \centering
    \begin{tabular}{c||cccc} 
     & \multicolumn{4}{c}{NC} \\ 
    NR  & $1$ & $2$ & $3$ & $4$ \\ \hline \hline
     $1$  & $130.1$ & $119.4^{*}$ & $116.0^{*}$ & $115.5^{*}$ \\ \hline
     $2$  & $121.0$ & $181.3$ & $135.0$ & $168.7$  \\ \hline
     $3$  & $120.0$ & $158.1$ & $165.9$ & $202.7$  \\ \hline
 $4$  & $120.2$ & $166.2$ & $236.2$ & $301.6$  \\ \hline
    \end{tabular}
    \label{Table:Average-Iterations-16}
\end{table}
\subsection{Computational Results} \label{subsection:Results}
\begin{figure}
        \centering
        \caption{$20 \times 20$ Hadamard matrix}
        \label{fig:Hadamard-matrix-20}
        \includegraphics[width=0.5\textwidth]{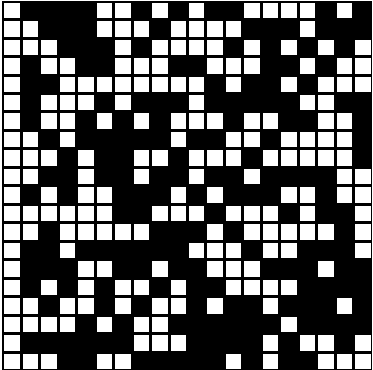}
    \end{figure}

\begin{figure}
        \centering
    \caption{$24 \times 24$ Hadamard matrix}
        \label{fig:Hadamard-matrix-24}
    \includegraphics[width=0.5\textwidth]{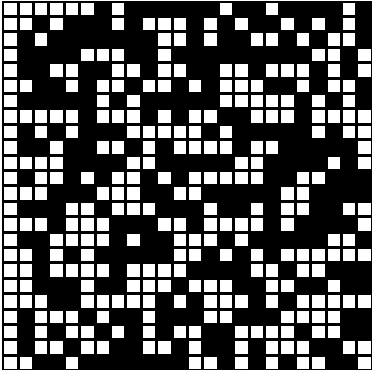}
    \end{figure}

\begin{figure}
        \centering
        \caption{$28 \times 28$ Hadamard matrix}
        \label{fig:Hadamard-matrix-28}  
    \includegraphics[width=0.5\textwidth]{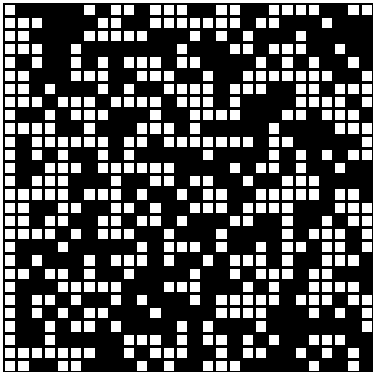}
    \end{figure}

\begin{figure}
    \centering
     \caption{$32 \times 32$ Hadamard matrix}\label{fig:Hadamard-matrix-32} \includegraphics[width=0.5\textwidth]{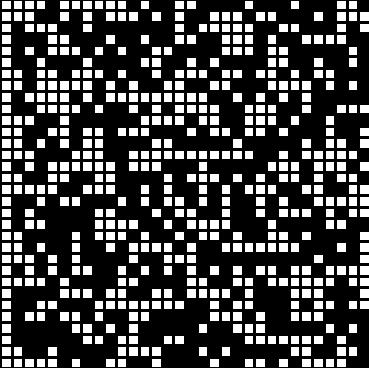}
    \end{figure}

The following are the Hadamard matrices found using  fitness function $F_{2}$ and the mutation kernel function \lstinline|Mutation|. 
\begin{enumerate}
    \item In Figure \ref{fig:Hadamard-matrix-20}, we show a $20 \times 20$ Hadamard matrix using the parameters $k=5$, $N=10^{3}$, $T=10^{7}$, $NC=2$, and $NR=8$. We obtained this result after $517$ seconds and $258875$ iterations.
    \item In Figure \ref{fig:Hadamard-matrix-24}, we show a $24 \times 24$ Hadamard matrix using the parameters $k=6$, $N=10^{3}$, $T=10^{7}$, $NC=2$, and $NR=10$. We obtained this result after $19997$ seconds and $9576814$ iterations.
    \item In Figure \ref{fig:Hadamard-matrix-28}, we show a $28 \times 28$ Hadamard matrix using the parameters $k=7$, $N=10^{4}$, $T=10^{8}$, $NC=1$, and $NR=10$. We obtained this result after $3054$ seconds and $428082$ iterations.
    \item In Figure \ref{fig:Hadamard-matrix-32}, we show a $32 \times 32$ Hadamard matrix using the parameters $k=8$, $N=2(10^{4})$, $T=10^{7}$, $NC=1$, and $NR=12$. We obtained this result after $94923$ seconds and $7472853$ iterations.
\end{enumerate}

\section{Conclusion}\label{conclusion}
In this paper, a genetic algorithm is proposed for the construction of Hadamard Matrices. The use of CuPy library in Python on graphics processing units enabled us to handle
populations of thousands of matrices and matrix operations in parallel. Our algorithm was able to find several Hadamard matrices up to size $32\times 32$. We did not observe any issues with the crossover part, but the effectiveness was sensitive to the mutation rate as it is typical in genetic algorithms. We also observed premature convergence rate of the fitness function with stalled convergence ultimately for large matrices.  
    
\section{Numerical Code}\label{code}
\begin{lstlisting}
import numpy as np
import cupy as cp 
import math
import os 

##### Raw kernel function: shuffling of +1/-1 in columns ####
Shuffle_Column = cp.RawKernel(r''' |\label{line:shuffle_begin}|
#include <curand_kernel.h> 
extern "C" __global__
void Shuffle_Column(char *Q, int seed, const int m, const int N)
{
    int k = blockDim.x*blockIdx.x+threadIdx.x; // matrix index
    int i = blockDim.y*blockIdx.y+threadIdx.y; // row index
    int j = blockDim.z*blockIdx.z+threadIdx.z; // column index
    unsigned long int seq, offset;
    int i1, i2;
    char ti1, ti2;
    seed=seed+k; seq = j;  offset = 0; 
    curandState h;     
    if ((k<4*N)&&(j>0)&&(j<m)&&(i==0)){
        curand_init(seed,seq,offset,&h);
        for(i1=0; i1<m-1; i1++){
            i2=(int)(curand_uniform(&h)*(m-i1)+i1);
            ti1=Q[m*m*k+m*i1+j]; 
            ti2=Q[m*m*k+m*i2+j];
            Q[m*m*k+m*i1+j]=ti2; 
            Q[m*m*k+m*i2+j]=ti1;
        }
    }
}
''', 'Shuffle_Column', backend='nvcc')  |\label{line:shuffle_end}|
##### End of shuffling kernel function ####


##### Raw kernel function: crossover ####
Crossover = cp.RawKernel(r''' |\label{line:crossover_begin}|
extern "C" __global__
void Crossover(char *Q, const int* c_cop, const int m, const int N)
{
    int k = blockDim.x*blockIdx.x+threadIdx.x;
    int i = blockDim.y*blockIdx.y+threadIdx.y; 
    int j = blockDim.z*blockIdx.z+threadIdx.z; 
    if (k<N){ 
        if((j<c_cop[k]) && (i<m)){
            Q[m*m*(k+2*N)+m*i+j]=Q[m*m*k+m*i+j]; //Copy the first part (columns before c_cop) of P1 into O1
            Q[m*m*(k+3*N)+m*i+j]=Q[m*m*(k+N)+m*i+j]; //Copy the first part of P2 into O2
       }
        if((j>=c_cop[k]) && (j<m) && (i<m)){
            Q[m*m*(k+3*N)+m*i+j]=Q[m*m*k+m*i+j]; //Copy the second part of P1 into O2
            Q[m*m*(k+2*N)+m*i+j]=Q[m*m*(k+N)+m*i+j]; //Copy the second part of P2 into O1
        }
    }
}
''', 'Crossover') |\label{line:crossover_end}|
##### End of crossover kernel function ####

##### Raw CUDA kernel function for Mutation ####
Mutation = cp.RawKernel(r''' |\label{line:mutation_begin}|
extern "C" __global__
void Mutation(char *Q, const int* row1, const int* row2, const int* col, const int NC, const int NR, const int m, const int N) 
{
    int k = blockDim.x*blockIdx.x+threadIdx.x;
    int i = blockDim.y*blockIdx.y+threadIdx.y;
    int j = blockDim.z*blockIdx.z+threadIdx.z;
    //k is the index of matrices, i is the index of rows, j is the index of columns
    int jc, ir, coljc, rowir1, rowir2;
    if ((k>=2*N) && (k<4*N)){ // Only the offspring are mutated
        if((i==0) && (j==0)){
            for(jc=0; jc<NC; jc++){
                coljc=col[(k-2*N)*NC+jc];
                for(ir=0; ir<NR; ir++){
                    rowir1=row1[(k-2*N)*NR+ir];
                    rowir2=row2[(k-2*N)*NR+ir];
                    if (Q[m*m*k+m*rowir1+coljc] != Q[m*m*k+m*rowir2+coljc]){ /* Only +1 and -1 pairs are flipped to keep the balance*/
                        Q[m*m*k+m*rowir1+coljc] *= -1;
                        Q[m*m*k+m*rowir2+coljc] *= -1;
                    }
                }
            }
        }
    }
}
''', 'Mutation') |\label{line:mutation_end}|
##### End of mutation kernel function ####

k=3
m=4*k # size of matrices
N=1000 # 4N is the population size
T=10**3 # maximum number of iterations
NC=10  # columns for swap  < m
NR=6  # rows for swap < m/2

blocksx = 8; blocksy = 8; blocksz = 8
blocks = (blocksx, blocksy, blocksz)
grids = (math.ceil(4*N/blocksx), math.ceil(m/blocksy), math.ceil(m/blocksz))

#### Initial Population ####
Pop = cp.ones((4*N, m, m), dtype=cp.int8)  |\label{line:popinit1}|
Pop[:,0:2*k,1:m] = -1  |\label{line:popinit2}|
seed=int.from_bytes(os.urandom(4), 'big') |\label{line:popinitseed}| 
Shuffle_Column(grids, blocks,(Pop, seed, m, N)) |\label{line:popinitcall}|

#Fitness Function
F=cp.count_nonzero(cp.matmul(cp.transpose(Pop,axes=(0,2,1)),Pop), axis=(1,2))-m |\label{line:fitness1}|
Fmin=cp.amin(F)

t=0
while (t<T) and (Fmin>0):
    #Selection
    #Moves best matrices (with the lowest norm) 
    # to the parent position at the beginning 
    # in random order
    Pop[0:2*N,:,:]=Pop[cp.random.permutation(cp.argsort(F)[:2*N]),:,:] |\label{line:selection}|
    
    #Crossover
    c_cop=cp.random.random_integers(1,m-1,N) |\label{line:crossoverpoint}|
    Crossover(grids, blocks, (Pop, c_cop, m, N)) |\label{line:crossover}|

    #Mutation
    rowindx1=cp.random.random_integers(0,m-1,size=(2*N,NR)) |\label{line:mutation1}|
    rowindx2=cp.random.random_integers(0,m-1,size=(2*N,NR))
    colindx=cp.random.random_integers(1,m-1,size=(2*N,NC))
    Mutation(grids, blocks, (Pop, rowindx1, rowindx2, colindx, NC, NR, m, N)) |\label{line:mutation2}|

    #Fitness Function
    F=cp.count_nonzero(cp.matmul(cp.transpose(Pop,axes=(0,2,1)),Pop), axis=(1,2))-m |\label{line:fitness2}|
\end{lstlisting}

\section*{Declarations}
The work of R. Ruiz was supported by a Dean’s Graduate Assistantship Award from the College of Sciences at the University of Texas Rio Grande Valley. The authors have no other relevant financial or non-financial interests or conflict of interests to disclose.

\bibliography{hadamard-bib}


\end{document}